\documentclass{elsart}

\usepackage{amssymb}
\usepackage{graphicx}

\begin{document}

\begin{frontmatter}


% use the corauthref command within \author for corru_npa3.texresponding author footnotes;

\title{\mbox{IBM-1} description of the fission products $^{108,110,112}$Ru}


\author[Leuv,Buc]{I. Stefanescu\corauthref{aut}},
\ead{Irina.Stefanescu$@$fys.kuleuven.be}
\corauth[aut]{Corresponding author.}
\author[Koeln]{A. Gelberg},
\author[Koeln]{J. Jolie},
\author[Caen]{P. Van Isacker},
\author[Koeln]{P. von Brentano},
%\author[Atlanta]{J. L. Wood}
\author[Nash,Berk]{Y. X. Luo},
\author[Thua,Nash]{S. J. Zhu},
\author[Berk]{J. O. Rasmussen},
\author[Nash]{J. H. Hamilton},
\author[Nash]{A. V. Ramayya},
\author[Thua]{X. L. Che}

\address[Leuv]{Instituut voor Kern- en Stralingsfysica, K.U. Leuven, Celestijnenlaan 200D,
B-3001 Leuven, Belgium}
\address[Koeln]{Institut f\"{u}r Kernphysik der Universit\"{a}t zu
K\"{o}ln, 50937 K\"{o}ln, Germany}
\address[Caen]{Grand Acc\'{e}l\'{e}rateur National d'Ions Lourds, CEA/DSM--CNRS/IN2P3, BP~55027
F-14076 Caen Cedex 5, France}
\address[Nash]{Physics Department, Vanderbilt University, Nashville,
TN 37235, USA}
\address[Berk]{Lawrence Berkeley National Lab., Berkeley, CA 94720
USA}
\address[Thua]{Department of Physics, Tsinghua University, Beijing,
100084, P. R. of China}
\address[Buc]{Horia-Hulubei National
Institute for Physics and Nuclear Engineering, PO-Box MG-6,
Bucharest, Romania}

\begin{abstract}
\mbox{IBM-1} calculations for the fission products
$^{108,110,112}$Ru have been carried out. The even-even isotopes
of Ru can be described as transitional nuclei situated between the
U(5) (spherical vibrator) and SO(6) ($\gamma$-unstable rotor)
symmetries of the Interacting Boson Model. At first, a Hamiltonian
with only one- and two-body terms has been used. Excitation
energies and $B$(E2) ratios of gamma transitions have been
calculated. A satisfactory agreement has been obtained, with the
exception of the odd-even staggering in the quasi-$\gamma$ bands
of $^{110,112}$Ru. The observed pattern is rather similar to the
one for a rigid triaxial rotor. A calculation based on a
Hamiltonian with three-body terms was able to remove this
discrepancy. The relation between the
IBM and the triaxial rotor model was also examined.
\end{abstract}
\end{frontmatter}
% main text

\section{Introduction}
\label{s_intro}
In recent years neutron-rich even-even isotopes of Ru
have been studied by gamma-ray spectroscopy.
In this work we will concentrate our attention
on the heavy fission products $^{108,110,112}$Ru,
for which new data are available.
In most cases, such nuclei were produced by spontaneous fission,
and the gamma-rays were studied
by using large Ge-detector arrays \cite{Hamilton95,Hamilton04,Shannon94}.
Recently, fission products with masses between about 100 and 112
were produced by a $^{252}$Cf source, and studied
with the Gammasphere array~\cite{Lu95,Che03,Hua03}.
In the most recent publication from the latter collaboration,
Zhu {\it et al.} ~\cite{Zhu06},
several new excited states and transitions have been reported.
Some neutron-rich even-even Ru isotopes were also produced
by the $\beta$-decay of Tc~\cite{Stachel84,Stachel82,Aysto90,Wang00}.
Recently, $^{104}$Ru was reinvestigated by Coulomb excitation~\cite{Sreb06}.

The impressive accumulation of experimental data during the last
5-6 years has created improved conditions for revisiting the
description of $^{108,110,112}$Ru by means of nuclear models. In
the paper by Stachel {\em et al.\/}~\cite{Stachel822} the authors
proposed to consider the isotopes between $^{98}$Ru and $^{110}$Ru
as belonging to a transition from the U(5) to the SO(6) limit of
the Interacting Boson Model (IBM)~\cite{Iachello87}. In
geometrical terms this would be equivalent to a transition between
a spherical vibrator~\cite{BM75} and a $\gamma$-unstable
rotor~\cite{WJ56}. The authors of~\cite{Stachel822} used a
schematic Hamiltonian, which can describe the main features of the
U(5) to SO(6) transition in even-even Ru isotopes.
In their conclusion they viewed this
simplified treatment only as a guideline. A more detailed approach
is necessary for a comparison of the data with the model for each
nucleus. Even-even Ru isotopes were studied in~\cite{Gianna95,Duarte98},
where the IBM-2 was used. In this variant of the model, separate proton and
neutron bosons are considered. The authors focussed their
attention on mixed-symmetry states~\cite{Arima77,Iachello84,Isacker86}.

More recently a search for a possible phase transition in these nuclei
was carried out in~\cite{Frank01}.
Again, a schematic Hamiltonian was used
to describe the U(5) to SO(6) transition.
The use of coherent states~\cite{Ginocchio80,Dieperink80}
allowed the authors to keep track of the dependence
on deformation of the total energy surface.
The authors came to the conclusion
that a phase transition takes place at $^{104}$Ru,
which can be considered as an example
of the E(5) critical symmetry
proposed by Iachello~\cite{Iachello00}.

Other models have also been used
for the calculation of excitation energies and transition strengths.
The Rotation-Vibration Model~\cite{Lu95,Eisenberg87}
and the Generalized Collective Model~\cite{Troltenier91,Troltenier96}
were applied to $^{108-112}$Ru.
A microscopically based Quadrupole Bohr Hamiltonian
was applied to $^{104}$Ru~\cite{Sreb06}.

Recently excited states in $^{109-112}$Ru
were populated in a $^{238}{\rm U}(\alpha ,{\rm f})$ fusion-fission reaction~\cite{Wu06}.
The bands above the backbending were interpreted
with the cranked shell model.

The main aim of this work is to describe the most important
observables of the heavy $^{108,110,112}$Ru isotopes such as
excitation energies, E2 branching ratios and the odd-even
staggering in the quasi-$\gamma$ bands by using the Interacting
Boson Model. We will use both the standard IBM-1 Hamiltonian and
an extended one, which also comprises a three-body term. We will
also discuss the relation between the  IBM-1 and the Rigid
Triaxial Rotor Model (RTRM). The calculations will be compared to
the latest experimental information presented in \cite{Zhu06}
which includes revisited $\gamma$-ray intensities and, for
$^{110,112}$Ru, a newly observed band-like structure built on the
proposed (4$^+_3$) state.

\section{The model}
\label{s_mod}
\subsection{The standard \mbox{IBM-1}}
\label{ss_sth} If we examine the systematics of excitation
energies in even-even $^{96-112}$Ru, we clearly see a transition
between two well-known patterns. In the light Ru isotopes, {\em
e.g.} $^{96,98}$Ru, the excitation energy ratio
$R_{4/2}=E(4_1^+)/E(2_1^+)\approx2$. With increasing neutron
number, $E(2_1^+)$ decreases, thus indicating an increase of
collectivity; the heaviest isotopes $^{110,112}$Ru have
$R_{4/2}\approx2.7$, a quite typical value for SO(6)-type nuclei.
In the light Ru isotopes we observe the two-phonon-triplet
$4^+,2^+,0^+$. Together with the energy ratio $R_{4/2}$, these are
characteristic features of a spherical vibrator. At the other
extreme, the $0^+_2$ state (if observed) is closer to the $3^+_1$
state, and it looks like belonging to a $6^+,4^+,3^+,0^+$
multiplet, as predicted by the Wilets-Jean model~\cite{WJ56} or
the SO(6) symmetry of the IBM. The energy ratios fit into the
picture too. This is a good reason for trying to describe the
even-even Ru isotopes in the framework of the algebraic
\mbox{IBM-1}.

The building blocks of the \mbox{IBM-1} are
$d$-bosons with an angular momentum 2,
and $s$-bosons, with an angular momentum 0. The corresponding creation
(annihilation) operators are $d^{\dag} (d)$ and $s^{\dag} (s)$,
respectively. A nucleus is characterized by the total boson number
$N$, which is equal to half the number of valence nucleons
(particles or holes). No distinction is made between proton
and neutron bosons.

A very important feature of the IBM is the conservation of the
total number of bosons. The operator
$\hat{N}=\hat{n}_d+\hat{n}_s$, where $\hat{n}_d$ and $\hat{n}_s$
are the $d$-boson and $s$-boson number operators, commutes with the
Hamiltonian and with the operators of electromagnetic transitions.
The boson number varies along the $^{98}$Ru to $^{112}$Ru chain
from $N=5$ to $N=10$.

It can be shown~\cite{Iachello87} that
the 36 linearly independent boson one-body operators
are the generators of  a Lie algebra
of the unitary group U(6).
A Casimir operator is defined by the condition
that it commutes with all elements of the algebra.

A fundamental concept of the IBM
is the existence of dynamical symmetries (DS),
which assumes the existence of a chain of subalgebras (subgroups)
starting from U(6) and ending at SO(3).
In a DS the Hamiltonian is a linear combination
of Casimir operators belonging to a chain of subalgebras.
We first consider the U(5) subalgebra chain
\begin{equation}
{\rm U}(6) \supset {\rm U}(5) \supset {\rm SO}(5) \supset {\rm SO}(3).
\label{e_u5}
\end{equation}
The quantum numbers of the chain~(\ref{e_u5}) are
the total number of bosons $N$,
the number of $d$-bosons $n_d$,
the $d$-boson seniority $\tau$
and the angular momentum $L$.
The main quantum number is $n_d$.
The label $\tau$, associated with SO(5),
here represents  the number of d-bosons
which are not coupled pairwise to angular momentum zero.
An additional quantum number $\nu_{\Delta}$
is necessary in order to specify completely the reduction from SO(5) to SO(3).
This quantum number is related to the number of triplets
of $d$-bosons coupled to angular momentum zero.

Besides the U(5) chain,
there are two other possibilities, namely SU(3) and SO(6).
We mentioned already that the heavy Ru isotopes
have properties which closely resemble those of the SO(6) symmetry.
The corresponding subalgebra chain is
\begin{equation}
{\rm U}(6) \supset {\rm SO}(6) \supset {\rm SO}(5) \supset {\rm SO}(3).
\label{e_so6}
\end{equation}
The algebra U(5) does not belong to this chain
and, as a consequence, the number of $d$-bosons $n_d$
is not a good quantum number.
On the other hand, due to the presence of SO(5),
the $d$-boson seniority $\tau$ is again a good quantum number.

The ``standard" \mbox{IBM-1} Hamiltonian contains
only one- and two-body operators.
The excitation spectrum of a single nucleus
(at fixed total boson number $N$)
depends on a maximum of six free parameters in the Hamiltonian.
A less general Hamiltonian
which is suitable for the description of the transition
between the U(5) and SO(6) DS is~\cite{Saha44}
\begin{equation}
\hat H=
\epsilon \hat n_d+
\kappa \hat Q^\chi\cdot\hat Q^\chi+
2\lambda \hat L\cdot\hat L+
\beta \hat n_d^2,
\label{e_sh}
\end{equation}
where $\hat Q^\chi$ is the quadrupole operator.
Its components are
\begin{equation}
\hat Q_\mu^\chi=
[d^\dag\times s + s^\dag\times\tilde d]^{(2)}_\mu
+\chi[d^\dag\times\tilde d]^{(2)}_\mu
\label{e_qop}
\end{equation}
where $\times$ denotes tensor coupling.
The effect of the $\beta\hat n_d^2$ term with $\beta<0$
is an increase of the moment of inertia
with increasing spin (or $\tau$).
This so-called ``$\tau$-compression" has been first used in~\cite{Pan92}.

The electric quadrupole transition operator is defined as
\begin{equation}
\hat T_\mu({\rm E2})=
q\left([d^\dag\times s + s^\dag\times\tilde d]^{(2)}_\mu
+\chi[d^\dag\times\tilde d]^{(2)}_\mu\right),
\end{equation}
where $q$ is an effective quadrupole charge.

In the {\em consistent-$Q$ formalism}~\cite{Casten90}
the parameter $\chi$ in $\hat T({\rm E2})$
has the same value as in the Hamiltonian.
If $\chi=0$, the E2 selection rule is $\Delta \tau=\pm1$;
if $\chi\neq0$, E2 transitions with $\Delta\tau=0,\pm2$ are also allowed.
For $\chi=0$, $\tau$ is a good quantum number
for the entire U(5) to SO(6) transition~\cite{Talmi93}.
In this case the wave functions have an important property.
For a given value of $\tau$,
$n_d$ can have only the values $\tau ,\tau +2,\tau+4$, etc.
In the U(5) limit this reduces to a fixed value of $n_d$.

The Hamiltonian~(\ref{e_sh}) displays a U(5) DS if $\kappa=0$
and, in this case, its excited states will be those of an anharmonic vibrator.
If, on the contrary, we want to obtain an SO(6) DS,
the parameters $\epsilon$, $\chi$ and $\beta$ must vanish.
If we examine the level schemes of the even-even Ru isotopes,
we notice that none of them is entirely consistent with either
the U(5) or the SO(6) symmetry.

In order to carry out a quantitative comparison of the data
with the theoretical predictions,
the Hamiltonian~(\ref{e_sh}) must be diagonalized numerically.
This has been done by using the program PHINT~\cite{Scholten82}.
The actual fitting procedure makes use of a fast graphical user
interface~\cite{Saha04}.
Its input contains the parameters used in eq.~(\ref{e_sh}).

As it will be shown in sect.~\ref{s_res},
most properties of the investigated nuclei
can be correctly described by the relatively simple Hamiltonian~(\ref{e_sh}).
However, the odd-even staggering
in the quasi-$\gamma$ bands of $^{110,112}$Ru
does not look like the pattern we should expect for SO(6)-type nuclei.
In this symmetry, in a first approximation,
the states belonging to a $\tau$-multiplet are degenerate.
If we switch on the term in $\hat L^2$, this degeneracy is lifted,
and the states with higher $L$ are raised in energy.
Since this contribution from SO(3) is small,
this will result in a slight lowering of the states with odd angular
momenta  with respect to those with even angular momenta.
This behaviour has been observed in most SO(6)-type nuclei,
including $^{108}$Ru.
There are a few SO(6) type nuclei in which this odd-even staggering
is replaced by a pattern which is closer to the RTRM.
In the latter case, for example, the $5^+_1$ state
is nearer to $4^+_2$ than to $6^+_2$, as it would be
required by the \mbox{IBM-1}~\cite{Casten85}
with only two-body terms in the Hamiltonian.
It has been observed that in the quasi-$\gamma$ bands of $^{110,112}$Ru,
the odd-even staggering is closer to the RTRM~\cite{Gore05}.
We will see in the next subsection
that this discrepancy can be removed
by use of an \mbox{IBM-1} Hamiltonian with cubic terms.

\subsection{The extended \mbox{IBM-1}}
\label{ss_exh}
The standard \mbox{IBM-1} Hamiltonian contains only
up to two-body interactions between the bosons.
Since the bosons are correlated objects with a complicated internal structure,
higher-order interactions between them are possible
and lead to an extended version of the model.
While a derivation of the form of such interactions
from microscopic considerations would be a highly non-trivial exercise,
the following argument
based on the coherent-state formalism~\cite{Ginocchio80,Dieperink80}
gives a phenomenological justification for them.
This formalism represents a bridge between algebraic and geometric models
and allows the construction of a potential in $\beta$ and $\gamma$
for any \mbox{IBM-1} Hamiltonian.
It can be shown~\cite{Ginocchio80,Dieperink80}
that the minimum of the potential
(which can be thought of as the equilibrium shape of the nucleus)
corresponding to a standard \mbox{IBM-1} Hamiltonian
occurs for $\gamma=0^\circ$ (prolate) or $\gamma=60^\circ$ (oblate).
Only with three-body interactions can the energy surface
exhibit a triaxial minimum~\cite{Piet81,Heyde84}
and hence such terms must be considered
if effects of triaxiality are to be included.

A general  \mbox{IBM-1} Hamiltonian with up to three-body
interactions can be diagonalized with the code ibm1~\cite{Pietxx}.
In the application to the Ru isotopes we use a two-body
Hamiltonian which is more general than in eq.~(\ref{e_sh}) and
reads
\begin{equation}
\hat H_{sd}^{(2)}= \epsilon' \hat n_d+ a_0\hat P^\dag\cdot\hat P+
a_1\hat L\cdot\hat L+ a_2\hat Q^\chi\cdot\hat Q^\chi+ a_3\hat
T_3\cdot\hat T_3+ a_4\hat T_4\cdot\hat T_4, \label{e_eh}
\end{equation}
where $\hat n_d$, $\hat L$ and $\hat Q^\chi$ have the same
definition as in sect.~\ref{ss_sth}, $\hat
T_\lambda=[d^\dag\times\tilde d]^{(\lambda)}$, and the SO(6)
pairing operator is defined as $\hat P^\dag=(s^\dag
s^\dag-d^\dag\cdot d^\dag)/2$. The simplified
Hamiltonian~(\ref{e_sh}) can be converted into the
form~(\ref{e_eh}) by noting the relation
\begin{equation}
\hat n_d^2= -5\hat n_d+ {\frac 2 9}\hat L\cdot\hat L+ {\frac 5
6}\hat T_3\cdot\hat T_3+ {\frac{35}{18}}\hat T_4\cdot\hat T_4.
\end{equation}
From this expression we can claim that, conversely, the general
Hamiltonian~(\ref{e_eh}) can be cast into the simplified
form~(\ref{e_sh}) if $a_0=0$ and $a_3/a_4=3/7$.

The general Hamiltonian with three-body terms contains many
interactions often involving $s$-bosons. The three-body
Hamiltonian used in the present calculation contains only
$d$-boson terms and can be written explicitly as
\begin{equation}
\hat H_d^{(3)}= \sum_J \tilde v_J [[d^\dag\times
d^\dag]^{(L)}\times d^\dag]^{(J)} \cdot[[\tilde d\times\tilde
d]^{(L')}\times\tilde d]^{(J)}. \label{e_h3}
\end{equation}
The allowed values of $J$ are 0, 2, 3, 4 and 6. For several $J$
more than one combination of intermediate angular momenta $L$ and
$L'$ is possible; these do not give rise to independent terms but
differ by a scale factor. To avoid the confusion caused by this
scale factor, we rewrite the Hamiltonian~(\ref{e_h3}) as
\begin{equation}
\hat H_d^{(3)}= \sum_J v_J B_J^\dag\cdot\tilde B_J, \qquad
B_J^\dag=N_{LJ}[[d^\dag\times d^\dag]^{(L)}\times d^\dag]^{(J)},
\end{equation}
where $N_{LJ}$ is defined such that $\langle
d^3;J|B_J^\dag\cdot\tilde B_J|d^3;J\rangle=1$, where
$|d^3;J\rangle$ is a normalized, symmetric state of three bosons
coupled to total angular momentum $J$. With this convention the
coefficients $v_J$ coincide with the expectation value of $\hat
H_d^{(3)}$ in the $|d^3;J\rangle$ state, $v_J=\langle d^3;J|\hat
H_d^{(3)}|d^3;J\rangle$.

The effect of three-body interactions~(\ref{e_h3}) was
investigated in~\cite{Heyde84}. It was found~\cite{Piet81,Heyde84}
that the three-body term with $J=3$ is most efficient for creating
a triaxial minimum in the energy surface. This procedure was
applied in~\cite{Casten85} to SO(6)-like Xe and Ba isotopes in the
mass region around $A=130$, as well as to $^{196}$Pt. These nuclei
display an unusual odd-even staggering in the quasi-$\gamma$ bands
with a pattern close to the RTRM. The experimental data could be
reproduced with the addition of a three-body term with $J=3$.

A more extensive investigation of the role played by the many
different three-body terms which have been ignored so far, is
highly desirable. This goes beyond the limited scope of this work
where only the term with $J=3$ in eq.~(\ref{e_h3}) has been
considered.

\section{Results and discussion}
\label{s_res}
\subsection{Fit with the standard Hamiltonian}
\label{ss_stf}
The IBM parameters were fitted to excitation
energies and $B$(E2) ratios as obtained in the recent
experiment \cite{Zhu06}. These data are characterized by better
statistics than obtained in previous experiments. The branching ratios
were determined by setting gates on transitions feeding the levels
of interest, thus reducing the possible systematic errors. The
$\tau$-forbidden transitions were very closely examined. These
high-quality experimental data constituted strict constraints on
the fitting procedure. In contrast, no constraints were
imposed on the parameters. For example, $\chi$ was not
automatically set to $\chi=0$ in $^{108}$Ru, in spite of its obvious
SO(6) character.

First, the standard version of the \mbox{IBM-1}
described in subsect.~\ref{ss_sth} was applied
to the study of the positive-parity bands in $^{108,110,112}$Ru.
The nuclei $^{108,112}$Ru have 20 valence nucleons or 10 bosons
relative to the nearest shell closures, $Z=50$ and $N=50,82$, respectively.
With $N=66$ neutrons, $^{110}$Ru
is located at the midshell between $N=50$ and $N=82$
and therefore has 22 valence particles or 11 bosons.
The parameters of the Hamiltonian~(\ref{e_sh})
were fitted to the experimental data in such a way
that the levels of the ground-state and quasi-$\gamma$ bands,
and some relevant $B$(E2) ratios are well reproduced.

\begin{figure*}
\vspace*{1cm}
\begin{center}
\includegraphics[height=10cm,width=14cm]{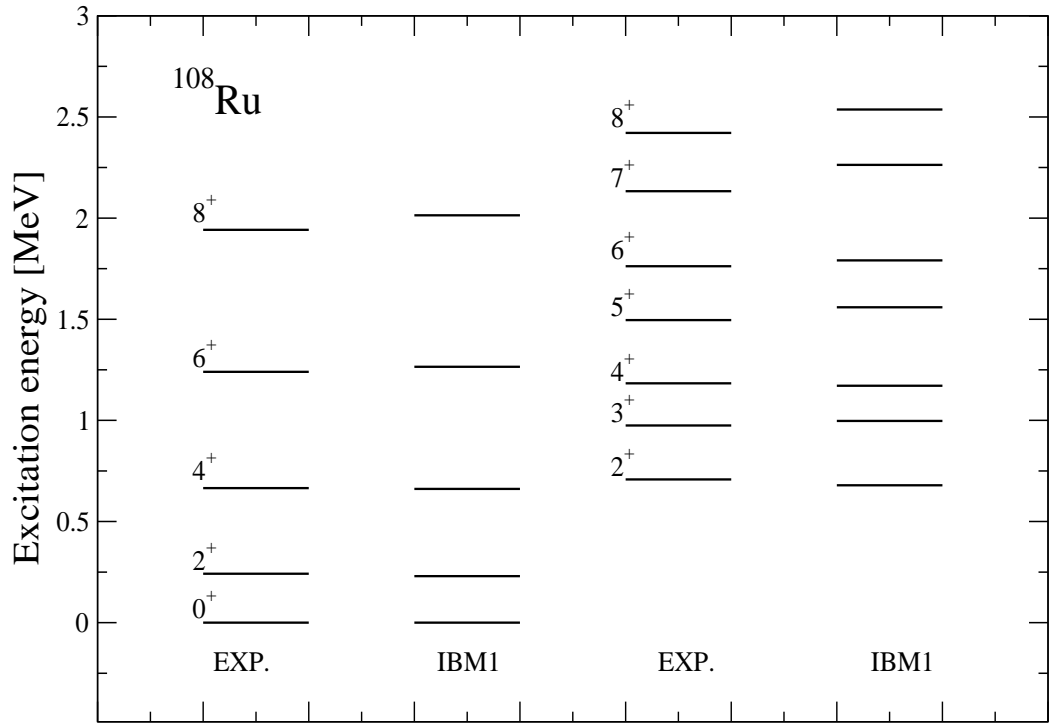}
\caption{Experimental and calculated level schemes for $^{108}$Ru.}
\label{f_ru108}
\end{center}
\end{figure*}

\begin{figure*}
\vspace*{1cm}
\begin{center}
\includegraphics[height=10cm,width=14cm]{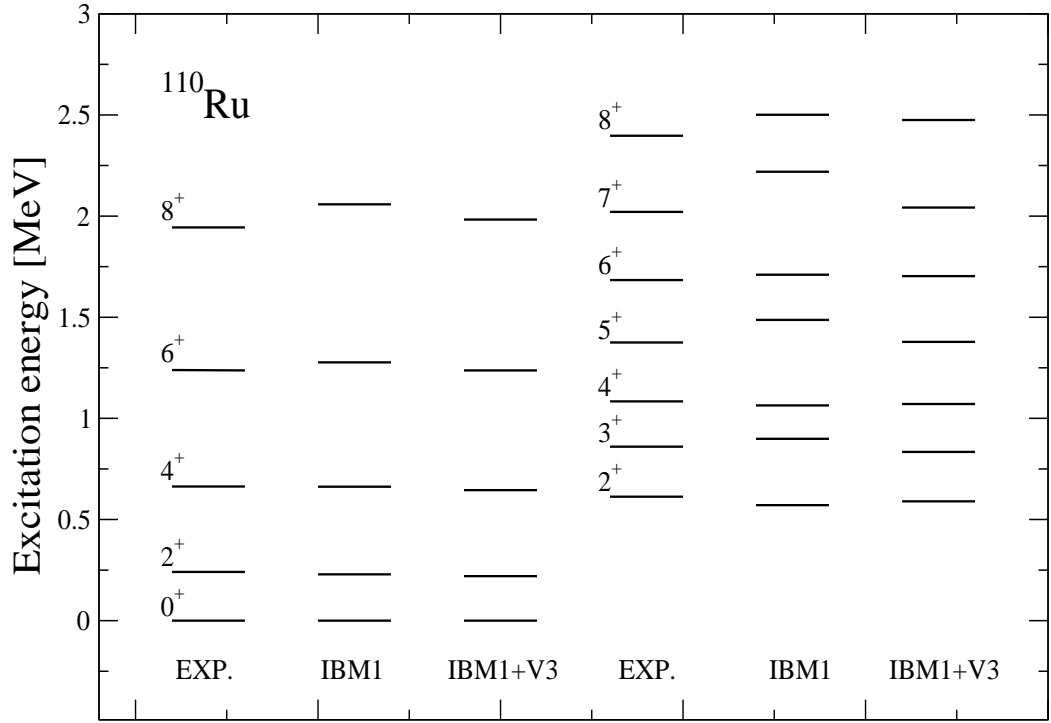}
\caption{Experimental and calculated level schemes for $^{110}$Ru.}
\label{f_ru110}
\end{center}
\end{figure*}

\begin{figure*}
\vspace*{1cm}
\begin{center}
\includegraphics[height=10cm,width=14cm]{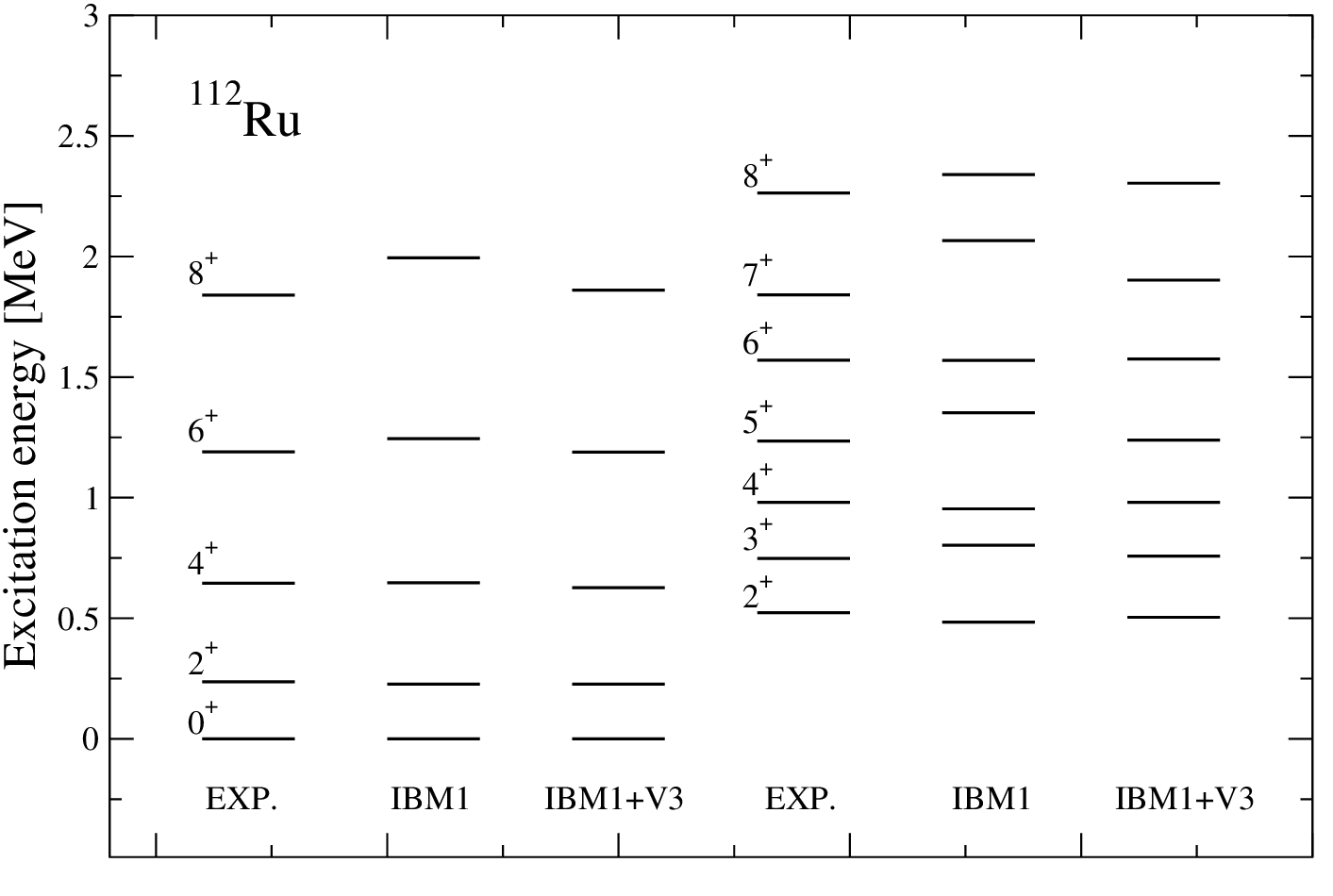}
\caption{Experimental and calculated level schemes for $^{112}$Ru.}
\label{f_ru112}
\end{center}
\end{figure*}

The comparison of the experimental levels
with those calculated with the standard \mbox{IBM-1}
is presented in figs.~\ref{f_ru108}, \ref{f_ru110} and \ref{f_ru112}.
For the ground-state bands only the levels up to spin $8^+$
were considered in the calculations
since above this spin value
the yrast bands exhibit a backbend in all nuclei.
Also in the quasi-$\gamma$ bands,
only levels up to spin $8^+$ were included in the fit.
The full set of parameters of the Hamiltonian~(\ref{e_sh})
obtained from the fit to the experimental data
is given in table~\ref{t_ibm}.
\begin{table*}
%\vspace*{1cm}
\caption{
\mbox{IBM-1} parameters used in the present calculations
with the standard Hamiltonian eq. (3).
All parameters are given in units of MeV,
except $\chi$ and $N$
which are dimensionless.}
\label{t_ibm}
\begin{center}
\begin{tabular}{cccc}
\hline
&$^{108}$Ru&$^{110}$Ru&$^{112}$Ru\\
\hline
$\epsilon$&$\phantom{-}0.804\phantom{0}$&$\phantom{-}0.713\phantom{0}$&$\phantom{-}0.509\phantom{0}$\\
$\chi$&$-0.128\phantom{0}$&$-0.077\phantom{0}$&$-0.060\phantom{0}$\\
$\kappa$&$-0.0485$&$-0.0466$&$-0.0439$\\
$\lambda$&$\phantom{-}0.014\phantom{0}$&$\phantom{-}0.015\phantom{0}$&$\phantom{-}0.016\phantom{0}$\\
$\beta$&$-0.106\phantom{0}$&$-0.091\phantom{0}$&$-0.089\phantom{0}$\\
$N$&10&11&10\\
\hline
\end{tabular}
\end{center}
\end{table*}
In the calculations all five parameters were allowed to vary.
They were found to change smoothly
when going from $^{108}$Ru to $^{112}$Ru,
indicating a slight change in structure when the boson number is varied.

A good agreement between the experimental
and calculated energy levels of the ground-state band
is obtained in all nuclei.
The quasi-$\gamma$ band is well reproduced
by the standard \mbox{IBM-1} only in $^{108}$Ru,
whereas the much weaker staggering observed in $^{110,112}$Ru
is clearly overestimated by the model
(see figs.~\ref{f_ru108}, \ref{f_ru110} and \ref{f_ru112}).
The calculated odd-even staggering is characteristic of the SO(6) limit,
whereas the staggering observed in $^{110,112}$Ru
is nearer to the RTRM pattern,
as discussed in the previous section.

\begin{figure*}
\vspace*{1cm}
\begin{center}
\includegraphics[height=8cm,width=10cm]{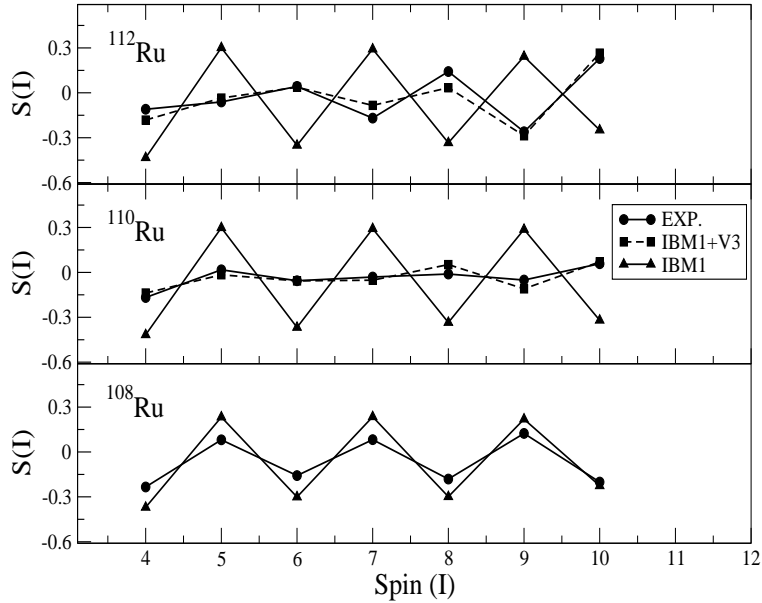}
\caption{ Experimental and calculated signature splitting for the
quasi-$\gamma$ bands in $^{108,110,112}$Ru.} \label{f_sig}
\end{center}
\end{figure*}

The accuracy of the fit can be easily seen
from the plot of the signature splitting
between the even- and odd-spin members of the band.
The signature splitting functions $S(I)$ for the experimental
and calculated levels in $^{108,110,112}$Ru
are represented in fig.~\ref{f_sig}
where $S(I)$ is given by~\cite{Zam}:
\begin{equation}
S(I)=
\frac{E(I)-E(I-1)}{E(I)-E(I-2)}\cdot
\frac{I(I+1)-(I-1)(I-2)}{I(I+1)-I(I-1)}-1.
\end{equation}
The $S(I)$ function vanishes
in the case of an axially symmetric rotor.
As it can be seen in the lowest panel of fig.~\ref{f_sig},
the experimental signature splitting in $^{108}$Ru
can be well described by the calculations.
In $^{110,112}$Ru, however, the model fails to reproduce
both the phase and magnitude of the experimental $S(I)$
(fig.~\ref{f_sig}, middle, top).
Furthermore, the model predicts in both nuclei
a slight decrease of the splitting with increasing spin,
in disagreement with the empirical observations.

\subsection{Fit with the extended Hamiltonian}
\label{ss_exf} Since the odd-even staggering pattern of the
quasi-$\gamma$ band in $^{110,112}$Ru resembles that of a
triaxially deformed rotor, it is natural to use the extended
\mbox{IBM-1} Hamiltonian, which may produce an energy surface with a
triaxial minimum. Adding the $J=3$ cubic term to the standard
\mbox{IBM-1} Hamiltonian lowers the $I^\pi=3^+,5^+,7^+,\dots$
states of the quasi-$\gamma$ band while the levels of the
ground-state band and the even-spin members of the quasi-$\gamma$
band are only slightly affected. Therefore, the starting
parameters for the fit with the extended version of the model were
those obtained in the simplified version (see table~\ref{t_ibm}),
converted to the standard representation~(\ref{e_eh}) (see
table~\ref{t_ibmc}, column ``IBM1''). The fitting procedure was
carried out for different values of the strength $v_3$ of the
cubic interaction until a reasonable agreement between the
experimental and calculated energy spacings of the quasi-$\gamma$
band was obtained. Finally, all parameters were readjusted in
order to obtain an overall agreement with the data.

\begin{table*}
\vspace*{1cm} \caption{ Parameters and rms deviations $\sigma$
given in units of MeV, except $\chi$ which is dimensionless.}
\label{t_ibmc}
\begin{center}
\begin{tabular}{cccccccc}
\hline
&$^{108}$Ru&&\multicolumn{2}{c}{$^{110}$Ru}&&\multicolumn{2}{c}{$^{112}$Ru}\\
&$N=10$&&\multicolumn{2}{c}{$N=11$}&&\multicolumn{2}{c}{$N=10$}\\
\cline{4-5}\cline{7-8}
&IBM1&&IBM1&IBM1+V3&&IBM1&IBM1+V3\\
\hline
$\epsilon'$&\phantom{--.}1.334&&\phantom{--.}1.168\phantom{0}&\phantom{--.}1.220\phantom{0}&&
\phantom{--.}1.035\phantom{0}&\phantom{--.}1.097\phantom{0}\\
$a_1$&$-0.010$&&$-0.0052$&$-0.0076$&&$-0.0038$&$-0.0075$\\
$a_2$&$-0.049$&&$-0.0466$&$-0.0466$&&$-0.0439$&$-0.0439$\\
$a_3$&$-0.088$&&$-0.076\phantom{0}$&$-0.065\phantom{0}$&&$-0.074\phantom{0}$&$-0.058\phantom{0}$\\
$a_4$&$-0.206$&&$-0.177\phantom{0}$&$-0.177\phantom{0}$&&$-0.173\phantom{0}$&$-0.173\phantom{0}$\\
$\chi$&$-0.128$&&$-0.077\phantom{0}$&$-0.127\phantom{0}$&&$-0.060\phantom{0}$&$-0.100\phantom{0}$\\
$v_3$&---&&---&$-0.030\phantom{0}$&&---&$-0.035\phantom{0}$\\
$\sigma_{\rm g}$&$\phantom{-}0.062$&&
$\phantom{-}0.086\phantom{0}$&$\phantom{-}0.030\phantom{0}$&&
$\phantom{-}0.095\phantom{0}$&$\phantom{-}0.025\phantom{0}$\\
$\sigma_\gamma$&$\phantom{-}0.072$&&
$\phantom{-}0.098\phantom{0}$&$\phantom{-}0.034\phantom{0}$&&
$\phantom{-}0.102\phantom{0}$&$\phantom{-}0.030\phantom{0}$\\
\hline
\end{tabular}
\end{center}
\end{table*}

The result of the fit with the extended \mbox{IBM-1} Hamiltonian
denoted IBM1+V3 is shown in fig.~\ref{f_ru110} for $^{110}$Ru and in
fig.~\ref{f_ru112} for $^{112}$Ru and the parameters used in the
calculations are given in table~\ref{t_ibmc}. The calculated
levels for the quasi-$\gamma$ band in both nuclei now show a more
regular energy spacing resulting in a much better agreement with
the data. The addition of the cubic term accounts very well for
the observed signature splitting
%and the slight increase with increasing spin
(see fig.~\ref{f_sig}). These observations can be made
quantitative by quoting the root-mean-square (rms) deviation
between calculated and observed excitation energies. This is done
in table~\ref{t_ibmc} where rms deviations are given for the
ground-state band up to $I^\pi=8^+$ (denoted as $\sigma_{\rm g}$) and for
the quasi-$\gamma$-band up to $I^\pi=8^+$ (denoted as
$\sigma_\gamma$). The quoted deviations prove that the addition of
a cubic interaction leads to a substantial improvement of the
energy fit for the quasi-$\gamma$ bands in $^{110}$Ru and
$^{112}$Ru without destroying the agreement for the ground-state
band.

A comparison between the experimental and calculated $B$(E2)
ratios for some relevant interband and intraband transitions is
given in table~\ref{t_branch}. The question may be asked
whether all examined transitions have E2 multipolarity, as it has
been assumed in most publications. The Evaluated Nuclear Structure
Data File (ENSDF) \cite{NNDC} and the Nuclear Data Sheets contain a few
values of the mixing ratio $\delta ({\rm E2/M1})$ for $^{108}$Ru. No
$\delta$ values are known for $^{110,112}$Ru. The relative $B$(E2)
values for the $2_2\rightarrow 2_1$ and $3_1\rightarrow 2_1$ transitions
in $^{108}$Ru have been corrected for M1 contributions with the factors
0.95 and 0.9, respectively.

In the proton-neutron IBM-2, M1 transitions between symmetric states
are forbidden. An M1 transition can occur only if there is an admixture
of states containing antisymmetric bosons pairs
($F$-spin mixing) \cite{Otsuka93}. Such admixtures
are usually weak in the excitation energy range considered in this work.

\begin{table*}
\vspace*{1cm} \caption{ Experimental and calculated E2 branching
ratios for $^{108,110,112}$Ru.} \label{t_branch}
\begin{center}
\begin{tabular}{ccccccccccc}
\hline &\multicolumn{2}{c}{$^{108}$Ru}&&
\multicolumn{3}{c}{$^{110}$Ru}& &\multicolumn{3}{c}{$^{112}$Ru}\\
\cline{2-3}\cline{5-7}\cline{9-11}
Ratio&EXP&IBM1&&EXP&IBM1&IBM1+V3&&EXP&IBM1&IBM1+V3\\
\hline
$\frac{2_2^+\rightarrow 2_1^+}{2_2^+\rightarrow 0_1^+}$&8.6(20)&7.1&&14.9(2)&12.1&9.3&&22.2(3)&10.6&16.7\\
$\frac{3_1^+\rightarrow 2_2^+}{3_1^+\rightarrow 2_1^+}$&15.9(14)&12.5&&20.4(3)&14.3&14.3&&21.7(4)&14.3&20.8\\
$\frac{4_2^+\rightarrow 2_2^+}{4_2^+\rightarrow 2_1^+}$&100(5)&100&&100(6)&10&500&&318(26)&166.7&1000\\
$\frac{4_2^+\rightarrow 2_2^+}{4_2^+\rightarrow 4_1^+}$&2.1(1)&2.4&&1.1(1)&1.9&1.6&&0.94(4)&1.85&1.3\\
$\frac{5_1^+\rightarrow 3_1^+}{5_1^+\rightarrow 4_1^+}$&10(1)&20&&25(1)&27.8&30.3&&37(2)&27.8&50\\
$\frac{4_3^+\rightarrow 3_1^+}{4_3^+\rightarrow 2_2^+}$&&&&2.75(20)&19.75&0.62&&0.66(3)&34.1&4.3\\
\hline
\end{tabular}
\end{center}
\end{table*}

Most calculations are in good agreement with the experimental
values. The small differences in the $B$(E2) ratios obtained with
the two versions of the model are due to the slightly different
values used for the $\chi$ parameter in the boson transition
operator $\hat T({\rm E2})$ (see tables~\ref{t_ibm}
and~\ref{t_ibmc}) and the inclusion of the cubic term in the
extended \mbox{IBM-1} Hamiltonian.

\subsection{Effective $\gamma$}
It is of interest to compare the calculation with the extended Hamiltonian
with results of a global calculation
of nuclear ground-state properties using the finite-range
liquid-drop model (FRLDM) in which deviations from axial symmetry
were allowed~\cite{Moeller06}. One of the regions in which nuclear
ground states display triaxial deformation comprises the heavy Ru
isotopes. In particular, the potential energy surface of
$^{108}$Ru has a shallow minimum at $\gamma\approx21^\circ$.
This energy surface can be described as being $\gamma$ soft.

\begin{figure*}
\vspace*{1cm}
\begin{center}
\includegraphics[width=10cm]{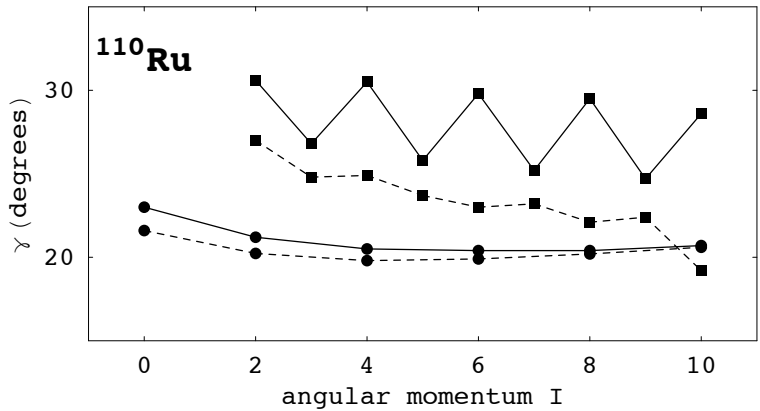}
\caption{Effective values of $\gamma_{\rm eff}$ for states belonging to the
ground (circles) and gamma (squares) bands in $^{110}$Ru. The full
(dashed) lines join the values of $\gamma_{\rm eff}$ obtained in the
standard (extended) \mbox{IBM-1} calculation.} \label{f_gameff}
\end{center}
\end{figure*}

Potentials obtained in the standard \mbox{IBM-1} are $\gamma$
unstable for $\chi=0$ in the quadrupole operator $\hat Q^\chi$.
For small, negative values of $\chi$ and an attractive quadrupole
interaction $\hat Q^\chi\cdot\hat Q^\chi$ a shallow minimum
develops for prolate deformation ($\gamma=0^\circ$). This minimum
shifts towards $\gamma=30^\circ$ if a cubic interaction $v_3$ is
added to the Hamiltonian. A more quantitative analysis can be made
by calculating the quadratic and cubic invariants and relating
them to the shape variables $\beta$ and $\gamma$ according to
\begin{equation}
\langle[\hat Q^\chi\times\hat Q^\chi]^{(0)}_0\rangle= \sqrt{\frac
1 5}\beta^2, \qquad \langle[\hat Q^\chi\times\hat Q^\chi\times\hat
Q^\chi]^{(0)}_0\rangle= -\sqrt{\frac{2}{35}}\beta^3\cos3\gamma,
\end{equation}
where $\langle\cdot\rangle$ denotes the expectation value for a
particular state. A value for $\gamma$ can thus be deduced from
the appropriate combination of the two
invariants~\cite{Elliott86}.
The deduced value can be considered as effective triaxiality,
denoted by $\gamma_{\rm eff}$,
as introduced by Yamazaki~\cite{Yamazaki78}.
Results obtained in this way for
$^{110}$Ru are shown in fig.~\ref{f_gameff}. (Similar results are
obtained for the two other isotopes.)
The values of $\gamma_{\rm eff}$ are fairly constant in the ground-state
band (except for a slightly higher value in the ground state) and
consistent with the results of the FRDLM for $^{108}$Ru. Also,
$\gamma_{\rm gsb}$ is found to be very close in the standard and
the extended \mbox{IBM-1}. There is, however, a significant
difference between the two calculations for the values of $\gamma_{\rm eff}$
for $\gamma$-band states with the pronounced even-odd staggering
effect in the standard \mbox{IBM-1} largely disappearing in the
extended \mbox{IBM-1}. The differences between the two
calculations are mainly due to the different choice for $\chi$
(see table~\ref{t_ibmc}) and the inclusion or not of a cubic
interaction.

\subsection{The 0$^+_2$ state}
\label{ss_zero}
Excited states at 976 and 1137~keV decaying to the $2^+_1$ level
were observed in previous works in $^{108,110}$Ru, ~\cite{Stachel84,Wang00}.
Spin and parity $0^+$ were assigned to these states based
on angular correlations measurements and the observed decay pattern.
The recent work of Zhu {\em et al.}~\cite{Zhu06}
confirms only the existence of a $0^+$ level at 975~keV in $^{108}$Ru.
States with $I^\pi=0^+$ located around 980~keV
were also observed in the lighter $^{104,106}$Ru isotopes.
Stachel {\em et al.} suggested that these state
might be based on intruder configurations~\cite{Stachel84,Stachel82}.
This configuration is produced by a proton(hole) pair excitation
accross the shell closure at Z=50.

Zamfir \cite{Zamfir02} has studied this phenomenon in $^{102}$Pd. A rather
peculiar behaviour of the $0^+_2$ state, which cannot be explained by either
IBM or E(5), has been ascribed to its intruder origin. At the same time, the
authors caution against extrapolating this feature to heavier Pd isotopes.
However, Lhersonneau and Wang \cite{Lhersonneau99,Wang01} investigated
$^{110,112,116}$Pd and observed typical intruder $0^+$ and $2^+$ states,
with an energy minimum at $^{110}$Pd. It is not yet clear whether the
same mechanism is working also in heavy Ru isotopes with $Z=44$.
In order to answer this question better experimental information
on the excited $0^+$ states in Ru is needed.

However, it is interesting to note that in $^{102}$Ru,
the $3^+_1$ state is located at 1522~keV
and that with increasing neutron number
the excitation energy of this state decreases,
reaching 975~keV in $^{108}$Ru.
The energies of $0^+_2$ states, however, remain around 970~keV
for the $^{102-108}$Ru nuclei.
In fact, this is the expected behaviour of the collective $0^+_2$ states
for the U(5) to SO(6) transition in \mbox{IBM-1}.
In the U(5) limit, the 0$^+_2$ and 3$^+_1$ states
have $n_d=2$ and 3, respectively,
whereas in the SO(6) limit both states belong to the multiplet with $\tau$=3.
The calculation with the standard version of the \mbox{IBM-1} Hamiltonian
and the set of parameters given in table~\ref{t_ibm}
predicts the 0$^+_2$ state in $^{108,110}$Ru
with $(\tau,\nu_\Delta)=(3,1)$ in the SO(6) limit
at an energy of 896 and 763~keV, respectively,
whereas the calculation for $^{110}$Ru with
the extended \mbox{IBM-1} Hamiltonian
predicts this state at 1002~keV.
The existence of $0^+$ states belonging to the $\sigma=N-2$ irrep of SO(6)
cannot be {\em a priori} excluded.
In any case, the observed $0^+_2$ state in $^{110}$Ru
does not fit well into the systematics.

\subsection{The band based on the $4^+_3$ state}
\label{ss_four}
\begin{figure*}
\vspace*{1cm}
\begin{center}
\includegraphics[height=10cm,width=14cm]{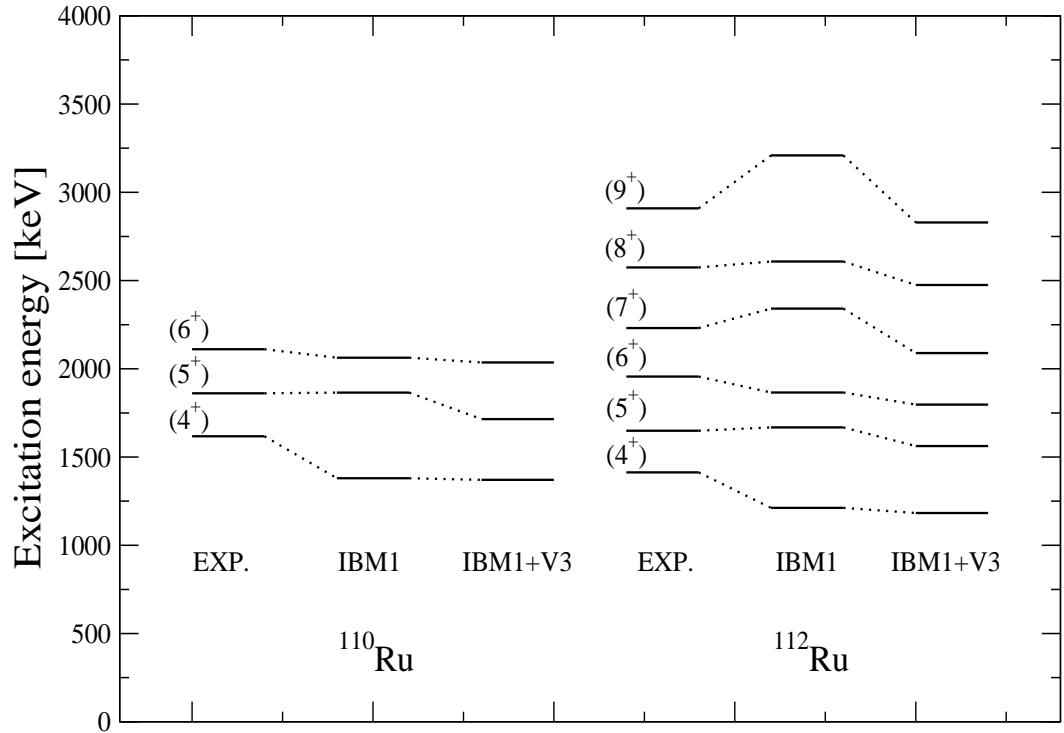}
\caption{ Experimental and calculated levels for the $K=4$ band in
$^{110,112}$Ru.} \label{f_k4}
\end{center}
\end{figure*}

An $I=4^+$ level located around 1.5~MeV excitation energy
and the band-like structure based on it
was identified in both $^{110}$Ru and $^{112}$Ru~\cite{Zhu06}.
This band was observed up to spin 6$^+$ in $^{110}$Ru
and 9$^+$ in $^{112}$Ru
and it was found to exhibit a rather regular energy spacing
between the even and odd spin members.
In the SO(6) limit of the standard \mbox{IBM-1}
such states have $\tau\ge4$, $\nu_\Delta=0$
and $\sigma=\sigma_{\rm max}$.
The results of the fits are presented in fig.~\ref{f_k4}.
The standard \mbox{IBM-1} Hamiltonian
fails to reproduce the observed weak staggering,
whereas the addition of the cubic term
improves  the quality of the agreement. However, both versions of the model
predict the $I=4^+$ bandhead
$\sim500$~keV lower in energy than experimentally observed.
The standard version of the model clearly overestimates
the ratio $B(E2;4^+_3\rightarrow3^+_1)/B(E2;4^+_3\rightarrow 2^+_2)$
since in the SO(6) limit of \mbox{IBM-1} such transitions
have $\Delta\tau$=1 and $\Delta\tau$=2, respectively.
The extended \mbox{IBM-1}
predicts a much lower branching ratio in both nuclei,
thus giving a slightly better agreement with the experimental observations
(see table~\ref{t_branch}).

In the rotational model such a $4^+$ state
is a candidate for a $\gamma$-$\gamma$ vibrational state,
generated by a double $\gamma$-phonon excitation~\cite{BM75}.
An equivalent interpretation can be formulated in the IBM
by means of the quadrupole ``phonon" excitation scheme~\cite{Siems94}.
In this case the quadrupole operator of eq.~(\ref{e_qop}) plays a role
similar to that of a phonon.
In this scheme excited states can be generated
by the multiple action of $\hat Q$ on the ground state.
For instance,
\begin{equation}
|2^+_1\rangle=A\hat Q|0^+_1\rangle,
\qquad
|2^+_2\rangle=B[\hat Q\times \hat Q]^{(2)}|0^+_1\rangle,
\end{equation}
and so on; $A$ and $B$ are normalization constants.
If we generate the $4^+_3$ state in this way,
we obtain in U(5) and SO(6) that
\begin{eqnarray}
|4^+_3 \rangle=\left(
a[[\hat Q\times\hat Q]^{(2)}\times
   [\hat Q\times\hat Q]^{(2)}]^{(4)}+
b [\hat Q\times\hat Q]^{(4)}\right)|0^+_1\rangle.
\end{eqnarray}
It has been shown~\cite{Jolos96} that $b\ll a$
so that, to a good approximation, the $4^+_3$ state is a four $Q$-phonon state.
It can be symbolically represented as
$[|2^+_2\rangle\otimes|2^+_2\rangle]^{(4)}$,
where the two states are coupled to $L=4$.
This is equivalent to what is usually called a double-$\gamma$ excitation.
As a general remark, the nature of the $4_3^+$ states is not yet
completely clear.

\subsection{IBM-1 versus RTRM}
\label{rtr}
The collective structure of $^{104,106,108,110,112}$Ru
was also previously discussed
\cite{Shannon94,Stachel84,Aysto90,Sreb06} in the framework of the
Rigid Triaxial Rotor Model (RTRM) \cite{Davydov58}. It has been
recognized that in these nuclei the $\gamma$ dependence of the
potential is intermediate between the two limiting cases of a
$\gamma$ unstable and a $\gamma$ rigid rotor
\cite{Shannon94,Stachel84}.

%From the comparison of the
%experimental and theoretical energy ratios $E_{2^+_2}/E_{2^+_1}$,
%values between $\gamma=$21$^\circ$ and 26$^\circ$ were deduced for
%$^{106-112}$Ru. These values were found to be only roughly
%consistent with those extracted from the $\gamma$-ray branching
%ratios of the $2^+_2$ levels \cite{Shannon94,Stachel84}.
In $^{108,110,112}$Ru, the experimental E2 branching ratios were
found to be in overall agreement with those predicted by the
RTRM calculations by using the $\gamma$-values 22.5$^\circ$,
24.2$^\circ$ and 26.4$^\circ$, respectively, as deduced from the
experimental $E_{2^+_2}/E_{2^+_1}$ energy ratio \cite{Shannon94}.

\begin{figure*}
\vspace*{1cm}
\begin{center}
\includegraphics[height=9cm,width=13cm]{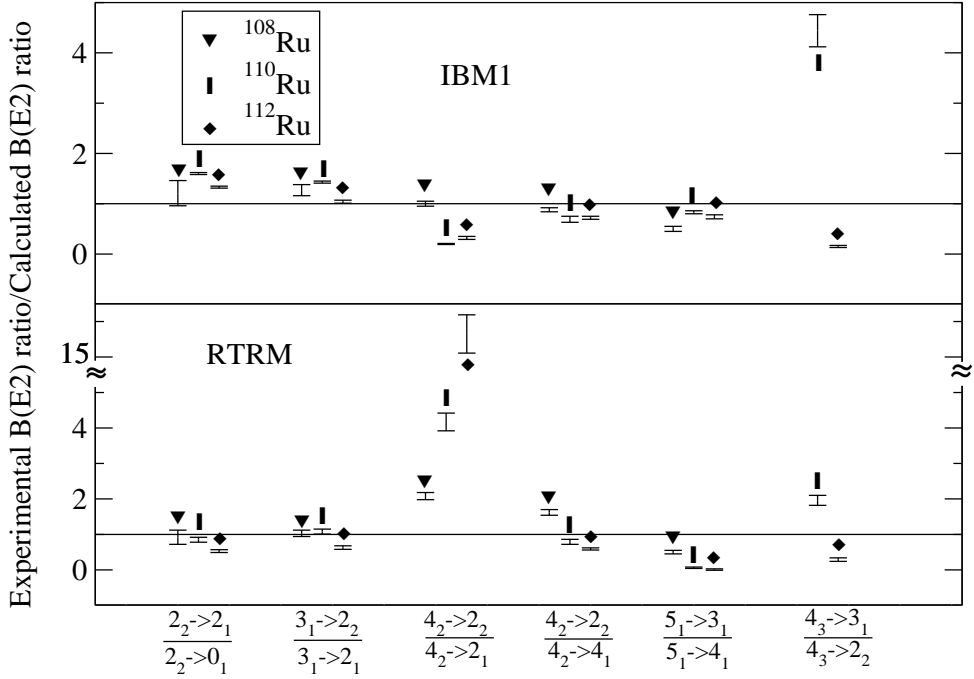}
\caption{ Experimental B(E2) ratios divided by the theoretical
ratios calculated with the standard ($^{108}$Ru) and extended
($^{110,112}$Ru) IBM-1 Hamiltonian (top) and RTR model (bottom)
with the $\gamma$-values given in ref. \cite{Shannon94}. Please
note the slight difference in the scales of the Y-axis for the two
plots.} \label{be2r}
\end{center}
\end{figure*}

\begin{figure*}
\vspace*{1cm}
\begin{center}
\includegraphics[height=8cm,width=9.5cm]{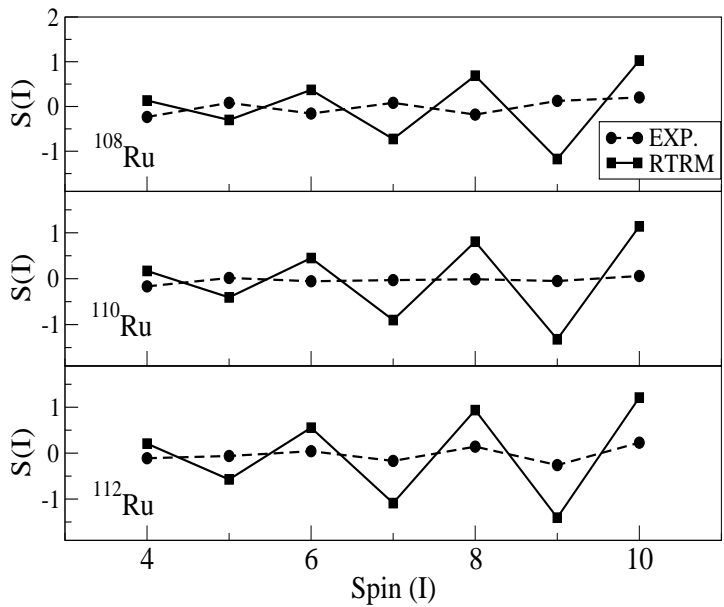}
\caption{Comparison between the experimental and calculated
signature splitting for the quasi-$\gamma$ bands in
$^{108,110,112}$Ru.}
\label{df}
\end{center}
\end{figure*}

Figure~\ref{be2r} shows the comparison between the experimental
B(E2) ratios divided by the ratios calculated with the
IBM1 Hamiltonian (top),
%IBM1 plus the cubic interaction (middle)
and RTRM model using the $\gamma$ values from \cite{Shannon94}
(bottom). In all cases, the experimental E2 branching ratios are
rather well described by the models. However, an RTRM calculation
of the odd-even staggering in the quasi-$\gamma$ band with the
values of the triaxiality parameter $\gamma$ from
\cite{Shannon94}, leads to a clear disagreement with experiment,
as seen in figure \ref{df}. This should be compared with the fits
shown in \mbox{fig. 4}.

The difficulty of correctly describing at the same time excitation
energies and E2 branching ratios is presumably the main problem of
the RTRM \cite{Shannon94,Wood04}. The authors of \cite{Wood04} assume
that the use of moments of inertia based on the irrotational flow
hypothesis is at the origin of the problem.

Contrary to the RTRM, the IBM, especially in the extended version
with 3-body terms in the Hamiltonian, is able to describe most of
the experimental data in the investigated nuclei (see figs.
\ref{f_ru110},\ref{f_ru112},\ref{f_sig},\ref{f_k4} and table
\ref{t_branch}).

Recently  triaxiality has been studied in a more general
geometrical context, by explicitly introducing a $\gamma$ soft
potential
 ~\cite{Fortunato04,Fortunato06}.

\subsection{Band structure}
\label{ss_band}
A sequence of states characterized by large values of $B$(E2)
for transitions inside the sequence
and connected by transitions with smaller $B$(E2)'s
to other sequences (inter-band transitions)
is usually called a {\em band}.
Another characteristic of a collective band
is the existence of a corresponding intrinsic state,
{\em i.e.} of a state defined in the intrinsic coordinate
frame (see~\cite{BM75}).

The concept of collective band has become so familiar
that level schemes are usually drawn in such a way
as to display the band structure.
The question arises whether the collective excited states of a nucleus
can always be classified into one-dimensional bands
inter-connected by weak E2 transitions.
This simple pattern cannot be seen in the investigated heavy Ru nuclides.
In a nucleus with clear band structure,
the ratio $B(E2;4_2\rightarrow 4_1)/B(E2;4_2\rightarrow 2_2)$
must be quite small.
In $^{108}$Ru this ratio has the value 0.69.
Both transitions in numerator and denominator have $\Delta\tau=-1$.
The ratio $B(E2;2_2\rightarrow 2_1)/B(E2;2_2\rightarrow 0_1)=15.6$ in the same nucleus.
In this case, the upper transition has $\Delta\tau=-1$,
while the lower one has $\Delta\tau=-2$ which is $\tau$-forbidden
both in SO(6) and in U(5) with $\chi=0$.
If the Alaga rule is applied, the ratio is 1.14.

What does the \mbox{IBM-1} predict?
In the SO(6) limit there are transitions with a large $B$(E2) value
between the band based upon the $2_2$ state and the gb.
Such bands have been called by Sakai~\cite{Sakai84} ``quasi-$\gamma$ bands".
The strong E2 transitions in the investigated nuclei
are those with $\Delta\tau=\pm1$.

It has been shown by Leviatan~\cite{Leviatan87}
that in the SO(6) limit of the IBM,
all states belonging to the lowest irrep of SO(6),
{\em i.e.} the states with $\sigma=N$,
can be projected out of a single intrinsic state.

\section{Conclusion}
\label{s_con}
Excitation energies and E2 ratios in $^{108,110,112}$Ru
have been calculated with the \mbox{IBM-1}.
First, a standard \mbox{IBM-1} approach
with only one- and two-body terms has been used.
A good fit was obtained for $^{108}$Ru.
However, the odd-even staggering in the quasi-$\gamma$ band of $^{110,112}$Ru
was not correctly described by the model.
The experimental staggering was rather close to the pattern
predicted by the triaxial rotor model.

Subsequently, an extended Hamiltonian
which contains three-body terms has been used.
It is known that this type of Hamiltonian
can generate a triaxial minimum of the total energy surface.
With this Hamiltonian the odd-even staggering
in the quasi-$\gamma$ band was correctly reproduced
while the excitation energies
and the $B$(E2) ratios of even-spin states were hardly affected.
%In other words, the three body-term acted as a perturbation.
In addition, effective values of $\gamma$ have been calculated.
%This extended Hamiltonian has
%been used for the first time for Ru isotopes.

It has been observed, in agreement with previous investigations,
that these nuclei are intermediate
between the U(5) and the SO(6) limits of the IBM.
There is no equivalent interpretation in the rigid triaxial rotor model.
However, none of these nuclides corresponds exactly to
either the U(5) or the SO(6) symmetry.
The $d$-boson seniority $\tau$ is an approximately good quantum number,
which allows one to classify the excited states
and to understand the $B$(E2) ratios.

The results obtained with the Interacting Boson Model
have been compared to those of the Rigid Triaxial Rotor Model.
While the $B$(E2) ratios have been equally well described by both models,
the odd-even staggering in the quasi-$\gamma$ bands
has been correctly reproduced only by the IBM.

The band structure of $^{108,110,112}$Ru has been discussed.
Contrary to the situation in the axially deformed nuclei,
the $B$(E2) ratios show that there are interband transitions
with E2 strengths of the same order of magnitude
as that of intraband transitions.

\section{Acknowledgements}

The authors would like to thank Dr. B. Saha, S. Heinze, Prof. P.
M\"{o}ller, Prof. V. Werner and Prof. J. Wood for interesting
discussions and communication of unpublished work. This work has
been supported by the Deutsche Forschungsgemeinschaft under grant
JO391/3-2, IAP Research Program no. P6/23 and FWO-Vlaanderen
(Belgium).

\end{document}